\documentclass[aps,pra,twocolumn,epsf,superscriptaddress]{revtex4}

\usepackage{amsmath}
\usepackage{graphicx}
\usepackage{dcolumn}
\usepackage{bm}

\newcommand{\hide}[1]{}

\newcommand{\eq}[1]{Eq.\,(\ref{#1})}
\newcommand{\eqs}[1]{Eqs.\,(\ref{#1})}
\newcommand{\noeq}[1]{(\ref{#1})}
\newcommand{\fig}[1]{Fig.\,\ref{#1}}

\newcommand{\tab}[1]{Table\,\ref{#1}}

\def\+#1{\ifmmode{{#1}^{\dagger}}\else{${#1}^{\dagger}$}\fi}
\def\bra#1{\ifmmode{\left< #1 \right|}\else{$\left< #1 \right|$}\fi}
\def\ket#1{\ifmmode{\left| #1 \right>}\else{$\left| #1 \right>$}\fi}
\def\braket#1#2{
  \ifmmode{\left< #1 | #2 \right>}
  \else{$\left< #1 | #2 \right>$}
  \fi
}
\def\half{\ifmmode{\frac 1 2}\else{${1\over2}$}\fi}
\def\mean#1{
  \ifmmode{\left< #1 \right>}
  \else{$\left< #1 \right>$}
  \fi
}

\newcommand{\Ec}{{\cal E}}
\newcommand{\Cc}{{\cal C}}
\newcommand{\Pc}{{\cal P}}
\newcommand{\Nc}{{\cal N}}
\newcommand{\e}{{\rm e}}
\newcommand{\ii}{{\rm i}}
\renewcommand{\i}{{\rm i}}
\newcommand{\tp}{{\tilde{p}}}
\newcommand{\ct}{{\check{\tau}}}
\newcommand{\s}{{\sigma}}
\newcommand{\Gm}{\Gamma}
\newcommand{\tGm}{\Gamma_f}
\newcommand{\gm}{\gamma}
\newcommand{\om}{\omega}
\newcommand{\Dl}{{\Delta}}

\newcommand{\cum}[1]{
  \left\langle\!\left\langle #1 \right\rangle\!\right\rangle}
\newcommand{\comm}[2]{\left[ #1 , #2 \right]}
\newcommand{\erw}[1]{\left\langle #1 \right\rangle}
\newcommand{\op}[2]{\left| #1 \right\rangle\left\langle #2 \right|}

\begin{document}

\title{Correlation in superradiance: A closed-form approach to cooperative effects}

\author{S.~F.~Yelin}
\affiliation{Department of Physics, University of Connecticut, Storrs, CT 06269}
\affiliation{ITAMP, Harvard-Smithsonian Center for Astrophysics,
                Cambridge, MA~02138}
\author{M. Ko\v{s}trun}
\affiliation{Department of Physics, University of Connecticut, Storrs, CT 06269}
\affiliation{ITAMP, Harvard-Smithsonian Center for Astrophysics,
                Cambridge, MA~02138}
\author{T.  Wang}
\affiliation{Department of Physics, University of Connecticut, Storrs, CT 06269}
\author{M.~Fleischhauer}
\affiliation{Universit\"at Kaiserslautern, Kaiserslautern, Germany}

\date{\today}

\begin{abstract}

We have developed a novel method to describe superradiance and related
cooperative and collective  effects in a closed form.
Using the method we derive a two-atom master 
equation in which any complexity of atomic levels, 
semiclassical coupling  fields and quantum fluctuations in the
fields can be included, at least in principle.
As an example, we consider the dynamics of an initially inverted
two-level system and show how even such in a simple system 
phenomena such as the initial radiation burst or broadening due to
dipole-dipole interactions occur, but it is also possible 
to estimate the population of the subradiant state during the radiative decay. Finally, we find that correlation only, not entanglement is responsible for superradiance.
\end{abstract}


\maketitle

\section{Introduction}

Following the appearance of Dicke's paper \cite{Dicke1953} on
spontaneous emission of radiation by an assembly of atoms,
cooperative effects such as superradiance were studied by 
many authors, e.g. \cite{RehlerEberly1971}.
Recently, there was a revived interest in, so called, Dicke states
because of their potential application in quantum information
processing~\cite{Fleischetal2000,FleischLukin2000,FleischLukin2002,Walsworthetal2002} and their importance in the behavior of
Bose-Einstein Condensates~\cite{MooreMeystre1999}. 

The main reasons to revisit superradiance in this article are (a) the need for an improved calculational model to support the revived interest, and (b) a set of novel phenomena that could be discovered using this model, such as subradiance and chirping, (c) the need for a simplified formalism to treat more complicated level systems, and (d) the improved understanding of the phenomena of collective vs. cooperative effects.

We have recently developed a novel method to study
collective effects in optically dense media \cite{FleischYelin1999:lorentz}.
Whereas collective
effects are caused by a radiative dipole-dipole interaction between
the atoms, cooperative effects also include 
the exchange of virtual photons between the atoms
that leads to the formation of Dicke states. 
The difficulty of calculating effects including atom-atom cooperation has
been the intractably large number of interconnected degrees of
freedom, even if just a few particles are involved. 
In this article we incorporate cooperative effects into
our formalism, leading to an effective two-atom master equation formalism for
superradiance. 
From the simplest version of this calculation we are
able to obtain good agreement with observed  signatures of
superradiance including the effects of dissipation and
the unique temporal build-up of a sharp flash of radiation.
In addition, our new formalism allows more complicated level structures, 
additional fields, and polarization to be easily considered as well.

What is the main difference between collective and cooperative effects? In this context, we define as ``collective effects'' all those, where a multi-particle ensemble responds differently than just the sum of all single-particle responses. Examples of collective effects in a atomic gas with radiative interactions include amplified spontaneous emission and radiation trapping (see e.g. \cite{FleischYelin1999:lorentz}). ``Cooperative effects,'' on the other hand, specifically arise from the exchange interaction between any two particles. The best-known example for this exchange interaction is the energy splitting between the symmetric and anti-symmetric two-atom superposition states, $(\ket{ab}\pm\ket{ba})/\sqrt{2}$, where $\ket{a}$ and $\ket{b}$ are the excited and ground states of a single two-level atom. In this article, we will quantify the exchange interaction as the non-diagonal two-atom term of a two-atom density matrix $\rho$, Tr{$\rho\op{ab}{ba}=\rho_{ab,ba}$}. We will show that the transition between non-cooperative and cooperative collective effects is given by the transition between $\rho_{ab,ba}=0$ tp $\rho_{ab,ba}\ne 0$. All typical signatures of superradiant behavior can be explained using this term.

Since the term $\rho_{ab,ba}$ describes the coherence between any two atoms, we will refer to it as ``correlation,'' and we will show (see, e.g., \fig{fig:c=10:all}) that the correlation is the reason for the build-up of superradiance.

\section{Model}

We consider an
arrangement of $N$ atoms which mutually interact via quantized electric
field in dipol and rotating-wave approximations (similar as was done in \cite{FleischYelin1999:lorentz}). 
We are interested in cooperative phenomena of the ensemble so we
distinguish two atoms in the arrangement, say $1$ and $2$, and
write the total Hamiltonian as
\begin{eqnarray}
  \label{eq:H}
  H &=&
  H_{field} + 
  \sum\limits_{i=1,2}
  \left\{
    H_0^i - 
    \vec{p}_i \left( \vec{E}_i + \Ec_i \right)
  \right\}\nonumber \\
&&  + \sum\limits_{j\ne 1,2} 
  \left\{ H_0^j - \vec{p}_j \left( \vec{E}_j + \Ec_j \right)
  \right\},
\end{eqnarray}
for the probe atoms $i=1,2$ and all other atoms $j\ne 1,2$.
Here, $H_0^{i,j}$ and $H_{field}$ are the free Hamiltonian of the $i,j$-th atom
at location $\vec r_{i,j}$, 
and that of the quantized radiation field, $\vec E$, respectively.
Furthermore, $\vec{E}_{i,j} = \vec{E}( \vec r_{i,j})$ is the quantized
field at the position of the $i,j$-th atom, while
$\Ec_j = \Ec( \vec r_j)$ is the external classical driving field.
$\vec{p}_{i,j} = \vec{p} (\vec r_{i,j})$ is the atom dipole operator.

\begin{figure}[ht]
\centerline{\includegraphics[width=0.85\linewidth]{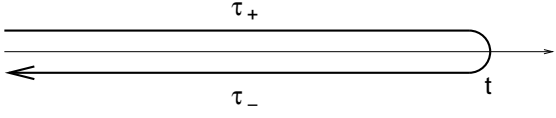}}
\label{fig:keldyshcontour}
\caption{The integration on the Schwinger-Keldysh contour follows the dark line from $-\infty$ via $t$ to $-\infty$. The time ordering is normal time ordering on the upper branch, inverse time ordering on the lower branch, and times on the lower come before times on the upper branch.}
\end{figure}

In the next step we move to an interaction picture using the time evolution
operator on the appropriately chosen Schwinger-Keldysh 
contour~\cite{Keldysh1964:contour}, shown in \fig{fig:keldyshcontour},
\begin{equation}
  \label{eq:keldyshcontour}
  S_\Cc = T_\Cc \exp
  \left\{
    \frac{\ii}{\hbar}\int_\Cc d\check{\tau} 
    \sum_{j=1,2}
    \vec{p}_j \left( \vec{E}_j + \Ec_j \right)
  \right\},
\end{equation}
where $T_\Cc$ is a time ordering operator on the
Schwinger-Keldysh contour, and is identical to standard time ordering 
operator $T$ on the upper or $+$ branch, and to its inverse $T^{-1}$ on
the lower or $-$ branch.
By this formal procedure each physical time $t$ is replaced by two
times on the contour, $t_+$ when going from $-\infty$ to $+\infty$ on the upper
or $+$ branch, and $t_-$ when going in opposite direction on the lower or
$-$ branch.
In this notation,
an infinitesimal complex quantity is added to what was previously real time $\tau$ in order to obtain $\check\tau$, 
where the sign of the infinitesimal complex part depends on
the branch of the Schwinger-Keldysh contour.

We find the expectation value of \eq{eq:keldyshcontour} by replacing formally 
\begin{equation}
\left\langle T_\Cc \exp{ \left\{\hat A\right\} } \right\rangle \;=\;
\exp{ \left\{ \sum_n\cum{T_\Cc\hat{A}^n} \right\} }\;,
\end{equation}
where $\hat A$ is the operator in the exponent of \eq{eq:keldyshcontour}. The cumulants contain the pure correlation, e.g., $\cum{\hat{x}}\equiv\erw{\hat{x}}$, $\cum{\hat{x}\hat{y}}=\erw{\hat{x}\hat{y}}-\erw{\hat{x}}\erw{\hat{y}}$, etc. Assuming that the radiation field is Gaussian allows us to bring to a
tractable form the cumulants involving the radiation field, which appear when
averaging the evolution 
operator~\cite{FleischYelin1999:lorentz,Gardiner1985,Fleisch1994:cumm}
with respect to the radiation field.
This told, in the effective evolution operator only terms up to quadratic
in electric field survive.
 
In addition,we separate positive and negative frequency components of the
atom dipole operator $p_\mu$ and the electric fields $E_\mu$ and $\Ec_\mu$,
\begin{eqnarray*}
    p_\mu(\check{\tau}) &=&
    p_\mu^+(\check{\tau})+p_\mu^-(\check{\tau}) \;=\;
    \tilde{p}_\mu^+(\check{\tau})\e^{-\ii\omega\tau} +
    \tilde{p}_\mu^-(\check{\tau}) \e^{\ii\omega\tau},\\
    E_\mu(\check{\tau}) &=&
    E_\mu^+(\check{\tau})+E_\mu^-(\check{\tau}),\\
    \tilde{E}_\mu^+(\check{\tau})\e^{-\ii\omega\tau} +
    \tilde{E}_\mu^-(\check{\tau}) \e^{\ii\omega\tau},\\
    \Ec_\mu(\check{\tau}) &=&
    \Ec_\mu^+(\check{\tau})+\Ec_\mu^-(\check{\tau}),
    \tilde{\Ec}_\mu^+(\check{\tau})\e^{-\ii\omega\tau} +
    \tilde{\Ec}_\mu^-(\check{\tau}) \e^{\ii\omega\tau},\\
\end{eqnarray*}
for the $\mu = x, y, z$--components along each of the coordinate axes. 
Thus, with a rotating wave approximation, \eq{eq:keldyshcontour} becomes
\begin{equation}  \label{eq:Seff}
S_\Cc^{\rm eff} \;=\;
T_\Cc\exp\left\{ \erw{\hat{A}^{(1)}}+\erw{\hat{A}^{(2)}}\right\}\;,
\end{equation}
where
\begin{eqnarray*}
\erw{\hat{A}^{(1)}} &=& \frac{\ii}{\hbar}\int\limits_\Cc d\check\tau \sum_{i=1,2} 
\left( \tilde{p}^+_{j\mu}(\check\tau)\Ec^-_{L,j\mu}(\check\tau) +
\tilde{p}^-_{j\mu}(\check\tau)\Ec^+_{L,j\mu}(\check\tau) \right)\\
\erw{\hat{A}^{(2)}} &=& -\frac{}{2\hbar^2}\int\limits_\Cc d\check\tau_1\int\limits_\Cc d\check\tau_2 \sum_{i,j=1,2}\\
&&\Big( \tilde{p}^+_{j\mu}(\check\tau_1)D_{i\mu,j\nu}(\check\tau_1,\check\tau_2)\tilde{p}^-_{j\nu}(\check\tau_2) + \\
&&    \tilde{p}^-_{j\mu}(\check\tau_1)C_{i\mu,j\nu}(\check\tau_1,\check\tau_2)\tilde{p}^+_{j\nu}(\check\tau_2) \Big)\;,
\end{eqnarray*}
where we used $\Ec_L = \Ec + \mean{\vec E}$ for the local field seen by
each of the two atoms.
The two cumulants that appear in \eq{eq:Seff} are
\begin{subequations}
  \label{eq:DC}
  \begin{eqnarray}
    D_{i\mu,j\nu}(\ct_1,\ct_2) &=&
    \cum{T_\Cc E_{i\mu}^-(\ct_1) E_{j\nu}^+(\ct_2)}
    \e^{-\ii\omega(\tau_1-\tau_2)} 
\\
    C_{i\mu,j\nu}(\ct_1,\ct_2) &=&
    \cum{T_\Cc E_{i\mu}^+(\ct_1) E_{j\nu}^-(\ct_2)}
    \e^{\ii\omega(\tau_1-\tau_2)} \;.
  \end{eqnarray}
\end{subequations}
The function $D$, for example, can then be expressed as
\begin{equation}
  \label{eq:d:full}
  D_{i\mu,j\nu}(\ct_1,\ct_2) = \e^{-\ii\omega(\tau_1-\tau_2)}
D_{i\mu,j\nu}^{\pm\pm}\;,
\end{equation}
with
\begin{equation} \nonumber
    \begin{array}{l@{\;\equiv\;}l}
      \cum{T\, E_\mu^-(\vec{r}_i,\tau_{1+}) 
        E_\nu^+(\vec{r}_j,\tau_{2+})} 
      & D_{\mu\nu}^{++},\\
      \cum{T^{-1}\, E_\mu^-(\vec{r}_i,\tau_{1-})
        E_\nu^+(\vec{r}_j,\tau_{2-})}
      &D_{\mu\nu}^{--}, \\
      \cum{E_\mu^-(\vec{r}_i,\tau_{1-}) E_\nu^+(\vec{r}_j,\tau_{2+}) }
      & D_{\mu\nu}^{-+},\\
      \cum{E_\nu^+(\vec{r}_j,\tau_{2-}) E_\mu^-(\vec{r}_i,\tau_{1+})}
      & D_{\mu\nu}^{+-},
    \end{array}
\end{equation}
where we used the properties of the time ordering operator $T_\Cc$ on
the Schwinger-Keldysh contour.
In the expression above a subscript sign ``+'' or ``-'' to the time
variable $\tau_{1,2}$ indicates the branch of the Keldysh contour on
which the time variable is located.
Similar expression can be written down for function $C$ as well.
We assume that there is only one type of atoms so we set
\begin{equation}
  \begin{array}{rcl}
    \tp_{\mu}^+(\vec r_i, t) & = & \wp_\mu \s_{i\mu}(t)\\
    \tp_{\mu}^-(\vec r_i, t)   & = & \wp_\mu \s_{i\mu}^\dagger(t),\\
  \end{array}
\end{equation}
where $\vec{\wp}$ is the dipole matrix element for all polarizations,
and $\hat{\vec{\s}}_i(t)$ is the dimensionless dipole vector operator of
individual atom $i$.
For real times, the effective time evolution operator \noeq{eq:Seff} becomes, after all the
contributions from the two branches are added together,
\begin{eqnarray}
  \label{eq:Seff:final}
    \lefteqn{S_\Cc^{\rm eff} \;=\;
    T_\Cc \exp \Bigg\{
      \frac{\ii\,\wp_\mu}{\hbar}
      \int\limits_{-\infty}^{+\infty} d\tau} \nonumber \\
    && \qquad
        \sum\limits_{i=1,2} \Big[
	\sum\limits_{\kappa=+,-}\/\kappa\,\s_{i\mu}(\tau_\kappa)\Ec^-_{L,i\mu}(\tau_\kappa) + {\rm H.c.} \Big]\nonumber\\
    && -
      \frac{\wp_\mu\wp_\nu}{2\hbar^2}
      \int\limits_{-\infty}^{+\infty}
      \int\limits_{-\infty}^{+\infty}
      d\tau_1\,
      d\tau_2\\
    && \qquad \sum\limits_{i,j=1,2} \Big[ \sum\limits_{\kappa,\lambda=+,-}
        \/\/\kappa\lambda\,
        \s_{i\mu}(\tau_{1\kappa})
        \big( D^{\kappa\lambda}_{i\mu,j\nu}(\tau_{1\kappa},\tau_{2\lambda})+
        \nonumber\\
    &&  \qquad\qquad C^{\kappa\lambda}_{i\mu,j\nu}(\tau_{1\kappa},\tau_{2\lambda}) \big)
        \s^+_{j\nu}(\tau_{2\lambda}) \Big] \Bigg\} \nonumber
\end{eqnarray}
Here, the superscripts in radiation field averages $C$ and $D$ determine
on which of the branches of the Schwinger-Keldysh contour respective time
argument lies, example of which was given in \eq{eq:d:full}.
Additionally, we assume that the field coherence
is shorter in duration than the 
coherence of the atomic operators $\s,\s^\dagger$, e.g., 
\begin{eqnarray}
\lefteqn{    \int d\tau_1 d\tau_2 
    \s(\tau_1) \cdot F(\tau_1,\tau_2) \cdot \s^\dagger(\tau_2) }\nonumber\\
    &=&\int d\tau \int d\tau'
    \s(\tau-\frac{\tau'}{2}) 
    F( \tau - \frac{\tau'}{2}, \tau +\frac{\tau'}{2} )
    \s^\dagger(\tau+\frac{\tau'}{2})\nonumber\\
    & \approx &
    \int d\tau
    \s(\tau)
    \left\{
    \int d\tau'
      F( \tau - \frac{\tau'}{2}, \tau +\frac{\tau'}{2} )
      \right\}
    \s^\dagger(\tau)\\
    &=&
    \int d\tau
    \s(\tau) 
    {\cal F}(\tau, \omega)
    \s^\dagger(\tau).\nonumber
\end{eqnarray}
This yields a more convenient way to reexpress the quantum correction
functions $C^{\kappa\lambda}$
and $D^{\kappa\lambda}$, with $\kappa,\lambda\in\{+,-\}$ of \eq{eq:Seff:final},
based on the following functions 
(see also~\cite{FleischYelin1999:lorentz} for details),
\begin{subequations}
  \label{eq:quantumcorrection}
  \begin{eqnarray}
    \lefteqn{\Gamma_{i\mu,j\nu}(\tau,\omega)  \;=\;
    \frac{\wp_\mu\wp_\nu}{\hbar^2} \int\limits_{-\infty}^{+\infty} d\tau' }
    \label{eq:quantumcorrection:Gamma}\\
    && \qquad\qquad \cum{E_{i\mu}^-(\tau)E_{j\nu}^+(\tau+\tau')}\e^{\ii\omega\tau'}, \nonumber \\
    \gamma_{i\mu,j\nu}(\tau,\omega) &=&
    \frac{\wp_\mu\wp_\nu}{\hbar^2}\int\limits_{-\infty}^{+\infty} d\tau' \\
    && \qquad \qquad \mean{\comm{E_{i\mu}^+(\tau)}{E_{j\nu}^-(\tau+\tau')}}
    \e^{\ii\omega\tau'},\nonumber \\
    H_{i\mu,j\nu}(\tau,\omega) &=&
    \frac{\hbar}{2\pi} \Pc\int\limits_{-\infty}^{+\infty} d\omega'
    \frac{\Gamma_{i\mu,j\nu}(\tau,\omega')}{\omega-\omega'},
    \label{eq:quantumcorrection:H}\\
    h_{i\mu,j\nu}(\tau,\omega) &=&
    \frac{\hbar}{2\pi} \Pc\int\limits_{-\infty}^{+\infty} d\omega'
    \frac{\gamma_{i\mu,j\nu}(\tau,\omega')}{\omega-\omega'}.
  \end{eqnarray}
\end{subequations}
Here, $\Pc$ denotes the principal value of the integral that follows it.
Please note $\Gamma$, \eq{eq:quantumcorrection:Gamma}, and $H$,
\eq{eq:quantumcorrection:H} are real and imaginery part of the Fourier transformed cumulant
$D^{-+}$, and so they are mutually related via the Kramers-Kronig relationship.

Term $\Gamma$ describes the decay and pump rates induced by the
incoherent photons inside the medium, while the
term $\gamma$ is the spontaneous ``down rate'' in the atomic medium. 
Term $H$ describes a collective light shift in addition to 
the inhomogeneous broadening, and is incorporated in our formalism.
Term $h$ is a spontaneous contribution to the light shift, the diagonal
terms of which amount to the Lamb shift.
As we are not interested in the Lamb shift, we consider these terms
either included in the free Hamiltonian $H_0$ (diagonal),
or zero (off-diagonal).
With the help of \eq{eq:quantumcorrection} we can write the effective two-atom density matrix
equation as
\begin{eqnarray}
  \label{eq:master}
\lefteqn{  \dot{\rho} \;=\;
  -\frac{\ii}{\hbar}\comm{H_0}{\rho} } \\
  &+& \frac{\ii}{\hbar}\sum\limits_{\mu} \wp_\mu \sum\limits_{j=1,2}
  \comm{\s_{j\mu}\Ec_{L,\mu}^-(\vec r_j) + 
    \s_{j\mu}^\dagger\Ec_{L\mu}^+(\vec r_j)}{\rho}\nonumber\\
  &+& \frac{\ii}{\hbar} \sum\limits_{\mu,\nu}\sum\limits_{j=1,2}
  H_{j\mu,j\nu} \comm{\comm{\s_{j\mu}}{\s_{j\nu}^\dagger}}{\rho}\nonumber \\
  &-& \sum\limits_{\mu,\nu}\sum\limits_{i,j=1,2} \frac{\Gamma_{i\mu,j\nu}}{2}
  \left(
    \comm{\rho\s_{i\mu}}{\s_{j\nu}^\dagger} 
    + \comm{\s_{i\mu}}{\s_{j\nu}^\dagger\rho} 
  \right)\nonumber\\
  &-&\sum\limits_{\mu,\nu}\sum\limits_{i,j=1,2}
  \frac{\Gamma_{i\mu,j\nu} + \gamma_{i\mu,j\nu}}{2}
  \left(
    \comm{\rho\s_{j\nu}^\dagger}{\s_{i\mu}} +
    \comm{\s_{j\nu}^\dagger}{\s_{i\mu}\rho}
  \right),\nonumber
\end{eqnarray}
where $\Ec_{L,\mu}(\vec r_j)$ is the local (Lorentz-Lorenz) field felt
by the $j$-th atom.
\eq{eq:master} is one of the main results of this paper. Up to this point we assumed a homogenous atomic medium, made a Markov approximation, and neglected correlations of higher than second order. All coherent fields are displayed in the term $\vec\Ec_L$, together with the first-order quantum corrections which take care of the local field. All second-order corrections are in the terms with $H$ and $\Gamma$. 
This equation looks similar to the effective master equation for large times,
derived in Refs.~\cite{Schwendimann1973,Bonifacioetal1971}. Note, however, that \eq{eq:master} is a two-atom equation! In addition, the second-order quantum correction terms, which have quite a different form in our case, contain most of the physics in the context of cooperative effects. It is also possible to find a closed-form expression for them, as will be seen in the next section.

\section{Cooperative effects in a homogeneous gas of two-level atoms}

For a gas of two-level atoms there is only one transition, and the coordinate indices $\mu$ and $\nu$ can be 
dropped; the effective two-atom two-level master
equation \noeq{eq:master} assumes the form
\begin{eqnarray}
  \label{eq:master:2level:hom}
    \dot{\rho} &=&
    -\frac{\ii}{\hbar}\comm{H_0}{\rho} +
    \sum\limits_{j=1,2}\frac{\ii}{\hbar}\wp 
    \comm{\s_j\Ec_{L}^-(\vec r_j) + \s_j^\dagger\Ec_{L}^+(\vec r_j)}{\rho}
  	\nonumber\\
    &&
    + \frac{\ii}{\hbar} \sum\limits_{j=1,2} H_{j,j}
    \comm{ \comm{\s_j}{\s_j^\dagger}}{\rho} \\
    &&
    - \half\sum\limits_{i,j=1,2} \Gamma_{i,j}
    \left( \comm{\rho\s_{i}}{\s_{j}^\dagger} +
      \comm{\s_{i}}{\s_{j}^\dagger\rho} \right)
	\nonumber \\
    &&
    -\half\sum\limits_{i,j=1,2} (\Gamma_{i,j}+\gamma_{i,j})
    \left( \comm{\rho\s_{j}^\dagger}{\s_{i}} +
      \comm{\s_{j}^\dagger}{\s_{i}\rho} \right). \nonumber
\end{eqnarray}
Let us introduce the notation for the two-atom density matrix as following,
\begin{equation}
  \label{eq:rho:2level}
  \rho_{\alpha\beta,\gamma\delta} \equiv
  \bra{\alpha_1\gamma_2} \rho \ket{\beta_1\delta_2}.
\end{equation}
Each of the two atoms can be in the ground state \ket{b}, or the excited
state \ket{a}, and \ket{\alpha_1,\gamma_2} is a product  
$\ket{\alpha}_1 \cdot \ket{\gamma}_2$ of the states of the first
and second atom.
The atomic lowering operator $\s$ assumes the form $\s=\op{b}{a}$. 
This allows us to introduce three real functions, namely $a,d$ and $n$, 
as follows
\begin{subequations}
  \label{eq:2level:newvariables}
  \begin{eqnarray}
    2\,a &=&
    2\rho_{aa,aa} +
    \rho_{aa,bb} +
    \rho_{bb,aa},\\
    d &=&
    \rho_{aa,bb} -
    \rho_{bb,aa},\\
    n &=& 
    \rho_{aa,aa} -
    \rho_{aa,bb} -
    \rho_{bb,aa} +
    \rho_{bb,bb}.
  \end{eqnarray}
\end{subequations}
Thus, $a(\vec r_1,\vec r_2;t)$ is the single-atom excited-state population averaged over both atoms, $d(\vec r_1,\vec r_2;t)$ is the difference in excited-state population for the two atoms, and $n(\vec r_1,\vec r_2;t)$ is the product of the inversions of the two atoms.
In new variables the equations of motion read
\begin{subequations}
  \label{eq:2level:sys}
  \begin{eqnarray}
    \dot{a} &=& -(2\Gamma_++\gamma) a + \Gamma_+ - \Gamma_- d \;,\\
    \dot{d} &=& -(2\Gamma_++\gamma) d - 2\Gamma_-(2a-1) \;,\\
    \dot{n} &=& -2(2\Gamma_++\gamma) n - 2\gamma(2a-1) \nonumber\\
        && + 
    4\left(\bar\Gamma_+ + \bar\Gamma_-\right)\rho_{ab,ba} +
    4\left(\bar\Gamma_+ - \bar\Gamma_-\right)\rho_{ba,ab}, \\
    \dot{\rho}_{ab,ba} &=&
    -\left( 
      \gamma + 2\Gamma_+ + \ii\Delta_{12} 
    \right)\,{\rho}_{ab,ba}  \\
    && + \left(
      \bar\Gamma_+ + \bar\Gamma_- \right) n .
  \end{eqnarray}
\end{subequations}
where we used
\begin{subequations}
  \label{eq:2level:parameters}
  \begin{eqnarray}
    2\Gamma_\pm &=& \Gamma_{11} \pm \Gamma_{22} \\
    2\bar{\Gamma}_\pm &=& \Gamma_{12} \pm \Gamma_{21} \\
    \Delta_{12} &=& \frac{2}{\hbar}(H_{11}-H_{22}).
  \end{eqnarray}
\end{subequations}
It will become clear, that the cross-coupling or correlation term $\rho_{ab,ba}$ carries the cooperative physics.

\subsection{Quantum corrections in a small sample approximation}

We further discuss the \eq{eq:2level:sys} in terms of small sample 
approximation, where we assume that all the coordinate dependence can be
dropped from the variables describing the system, $a$, $n$, $d$, and 
$\rho_{ab,ab}$.
That is,  we neglect retardation effects of propagation
of the electromagnetic field through the sample, that is, all time changes
in the atomic variables propagate instantaneously through the sample. This is a good approximation as long as the ``cooperative'' time $\tau_C=(\Nc_{\rm exc}\gamma\mu)^{-1}$, with $\Nc_{\rm exc}$ the density of atoms in the excited state, $\gamma$ the vacuum decay, and $\mu$ a geometric factor, is longer than the maximum propagation time \footnote{$\tau_C$ gives a good estimate for the shortest possible superradiance timescales. The actual superradiance time $\tau_{\rm sr}$, i.e., the time for the buildup of radiation, is the one that finally has to be short compared to the propagation time. Ususally, $\tau_{\rm sr}>\tau_C$}.

In the small sample approximation, thus, there is no difference between the
atoms 1 and 2 with respect to their decay rates.
That is, in \eq{eq:2level:sys} we set
$\Gm_+ = \Gm_{11} = \Gm_{22} \equiv \Gamma$.
Similarly, there is no spatial difference between the decay rates
so 
$\Gamma_- = \bar\Gamma_- = 0$.
For other decay rates we have
$\bar\Gamma_+ = \Gm_{12} =  \Gm_{21} \equiv \bar\Gamma$.
Finally, because in a homogeneous sample the result is invariant with respect
to exchange of the indices of two atoms, we have $\Dl_{12} =0$.
This said, the system \noeq{eq:2level:sys} reduces to,
\begin{subequations}
  \label{eq:ss:sys}
  \begin{eqnarray}
    \dot{a} &=& -(2\Gamma+\gamma) a + \Gamma\;,\\
    \dot{n} &=& -2(2\Gamma+\gamma) n - 2\gamma(2a-1) 
    + 8\bar{\Gamma} \rho_{ab,ba}, \\
    \dot{\rho}_{ab,ba} &=& - (2\Gamma+\gamma) \rho_{ab,ba} + \bar{\Gamma} n.
  \end{eqnarray}
\end{subequations}


Let us now discuss the decay matrices $\Gamma_{i,j}$, where
$i,j = 1,2$, and which we introduced in \eq{eq:quantumcorrection},
in greater detail.
Our discussion is based on the formalism for calculation the quantum
corrections to the electric field in a responsive media, 
that we have previously developed in~\cite{FleischYelin1999:lorentz}.
We need the exact Green's function for the Maxwell 
field $E$,
\begin{equation}
  \label{eq:qc:d}
  D_{\mu\nu}(\check{1},\check{2}) =
  \cum{T_\Cc E_\mu^-(\check{1}) E_\nu^+(\check{2})},
\end{equation}
where $E_0$ would be the field without medium present,
as calculated along the Schwinger-Keldysh contour. Here, we abbreviate $\vec r_1, \check{t}_1$ by $\check 1$, etc.
The Green's function for the free field $E_0$ is given by
\begin{equation}
  \label{eq:qc:d0}
  D_{0\mu\nu}(\check{1},\check{2}) =
  \cum{T_\Cc E_{0\mu}^-(\check{1})
    E_{0\nu}^+(\check{2})}.
\end{equation}
The contour Green's function $D$ contains four real-time functions,
depending where on the Keldysh-Schwinger contour the two time points
are located,
$D^{++}$, $D^{-+}$, $D^{+-}$, and $D^{--}$,
as discussed in \cite{FetterWalecka1971}.
We find approximately \cite{FleischYelin1999:lorentz}, $D_0^{++}\approx D^{\rm adv}$, 
$D_0^{-+}\approx 0$, $D_0^{+-}\approx D_0^{\rm adv}-D_0^{\rm ret}$, 
and $D_0^{--}\approx -D_0^{\rm ret}$, 
where $D^{\rm ret}$ and $D^{\rm adv}$ are
retarded and advanced Green's functions, respectively.
A solution to the atom-field interaction can be written  in terms
of a Dyson equation,
\begin{eqnarray}
  \label{eq:qc:d:pm}
\lefteqn{  D_{\alpha\beta}^{-+}(1,2) =
  -\iint\limits_{-\infty}^{+\infty} dt_1' dt_2' }\\
&&  \iint_{V_1V_2} d\vec{r'}^3_1 d\vec{r'}^3_2 D^{\rm ret}_{\alpha\mu}(1,1')
  P^{\rm S}_{\mu\nu}(1',2')D_{\nu\beta}^{\rm adv}(2',2), \nonumber
\end{eqnarray}
where
\begin{eqnarray}
  \label{eq:qc:d:ret}
\lefteqn{  D_{\alpha\beta}^{\rm ret}(1,2) =
  D_{0\alpha\beta}^{\rm ret}(1,2)-
  \iint\limits_{-\infty}^{+\infty} dt_1'dt_2'} \\
  &&\iint_{V_1V_2} d\vec{r}_1'd\vec{r}_2' 
  D_{0\alpha\mu}^{\rm ret}(1,1') P_{\mu\nu}^{\rm ret}(1',2') 
  D_{\nu\beta}^{\rm adv}(2',2),\nonumber
\end{eqnarray}
with $D_{\alpha\beta}^{\rm adv}(1,2) = D_{\beta\alpha}^{\rm ret}(2,1)$.
In \eqs{eq:qc:d:pm} and \noeq{eq:qc:d:ret} figure two source (polarization)
functions.
In the lowest order in the atom-field
coupling, the polarization function is given by a correlation function of
dipole operators of noninteracting atoms,
\begin{equation}
  \label{eq:polarization:s}
  P^{(1,2)\text{s}}(\vec{r}_i,\vec{r}_j;t_1,t_2) 
  = \frac{\wp^2}{\hbar^2} \Nc^{(1,2)} \cum{\s_i^\dagger(t_1)\s_j(t_2)},
\end{equation}
and for retarded polarization,
\begin{eqnarray}
  \label{eq:polarization:ret}
  \lefteqn{P^{(1,2)\text{ret}}(\vec{r}_i,\vec{r}_j;t_1,t_2) =}\\
&&  \frac{\wp^2}{\hbar^2}\Nc^{(1,2)}\Theta(t_1-t_2)
  \erw{\comm{\s_i^\dagger(t_1)}{\s_j(t_2)}}.\nonumber
\end{eqnarray}
In \eqs{eq:polarization:s} and \noeq{eq:polarization:ret} the superscript
$1$ stands for a one-atom source function, i.e., $i=j$,
while $2$ is for a two-atom source function, and $i\ne j$.
The cumulants can be found by using the quantum-regression theorem
and a Laplace transformation, as done below.

The differential equation for the single atom density-matrix cross-term,
$\rho_{ab}$, is found by tracing over the second atom,
\begin{equation}
  \label{eq:avgpolarization}
  \dot \rho_{ab} = \dot\rho_{ab,aa}+\dot\rho_{ab,bb} = -\left(
    \frac{\gamma}2 +\Gamma + i \Delta \right) \, \rho_{ab}
\end{equation}
with $\Delta = \Delta_i\equiv 2 H_{ii}/\hbar$. Upon comparison with the same equation for a dilute gas, we find that the second-order correction $H_{ii}$ in this case takes on the role of a detuning from resonance. Since, in general, $H_{ii}$, along with $\Gamma$, changes over time, a chirp of the radiated light can be expected.

We can Laplace-transform the source functions from $\tau$ to $\lambda$ with $t_1\rightarrow t+\tau$ and $t_2\rightarrow t$ for $t_1>t_2$ (and vice versa).
Thus we find for \eq{eq:avgpolarization} (and its complex conjugate)
\begin{equation}
  \bar{\rho}_{ab} = 
  \frac{\rho_{ab}(\tau=0)}{\lambda+\gamma/2+\Gamma+\i\Delta},
  \quad
  \bar{\rho}_{ba} =
  \frac{\rho_{ba}(\tau=0)}{\lambda+\gamma/2+\Gamma-\i\Delta},
\end{equation}
Here
$\mean{\bar{\sigma}_i(\lambda,t)}=\bar{\rho}_{ab}$. Using these solutions and the quantum-regression theorem \cite{meystre}, we find for the Laplace-transformed single-atom cumulants ($i=j$)
\begin{equation}
  \label{eq:sigcum:1}
  \begin{array}{lclcl}
    \cum{\bar{\s}^\dagger_i(\lambda,t)\s_i(t)} &=&
    \frac{a(t)}{\lambda+\gamma/2+\Gamma-\i\Delta} &=&
    \cum{\s^\dagger_i(t)\bar{\s}_i(\lambda,t)}^*,\\
    \cum{\bar\s_i(\lambda,t)\s_i^\dagger(t)} &=&
    \frac{1-a(t)}{\lambda+\gamma/2+\Gamma+\i\Delta} &=&
    \cum{\s_i(t)\bar\s_i^\dagger(\lambda,t)}^*,
  \end{array}
\end{equation}
and for the two-atom cumulants ($i\ne j$)
\begin{equation}
  \label{eq:sigcum:2}
  \begin{array}{lclcl}
    \cum{\bar\s_i^\dagger(\lambda,t)\s_j(t)} &=&
    \frac{x(t)}{\lambda+\gamma/2+\Gamma-\i\Delta} &=&
    \cum{\s_i^\dagger(t)\bar\s_j(\lambda,t)}^*, \\
    \cum{\bar\s_j(\lambda,t)\s_i^\dagger(t)} &=& 
    \frac{x(t)}{\lambda+\gamma/2+\Gamma+\i\Delta}&=&
    \cum{\s_j(t)\bar\s_i^\dagger(\lambda,t)}^*.
  \end{array}
\end{equation}
In a simple case like this we can go from Laplace- to Fourier-transform by just replacing $\lambda$ by $\i\Delta'$, where $\Delta'$ is a frequency.
Then, we find for the source functions in Fourier space
\begin{eqnarray}
P^{(1)\text{s}}(\vec x,\vec r;\Delta,t)&=&\frac{\wp^2}{\hbar^2}\Nc\frac{2\,a(\vec x,t)(\gamma/2+\Gm)}{(\gm/2+\Gm)^2+(\Dl'-\Dl)^2}\\
P^{(2)\text{s}}(\vec x, \vec r;\Delta,t)&=&\frac{\wp^2}{\hbar^2}\Nc^2\frac{2\,x(\vec x,t)(\gamma/2+\Gm)}{(\gm/2+\Gm)^2+(\Dl'-\Dl)^2}\\
P^{(1)\text{ret}}(\vec x, \vec r;\Delta,t)&=&\frac{\wp^2}{\hbar^2}\Nc\frac{1-2\,a(\vec x,t)}{\gm/2+\Gm+\ii(\Dl'-\Dl)},
\end{eqnarray}
where $\vec r_i-\vec r_j=\vec x$ and $\vec r_j=\vec r$. Since in this case there is no external light field present, we can set $\Delta'=0$.

The equations above allow us to write down closed form expressions for
both $\Gamma$ and $\bar\Gamma$, where
\begin{subequations}
  \label{eq:GammaGammabar}
  \begin{eqnarray}
    \lefteqn{\Gm(\Dl,t) \,=\,
    \frac{\wp^2}{\hbar^2}\tilde{D}^{-+}(\vec{r}_1=\vec{r}_2,\Dl,t) \;=}\\
    &=&
    \frac{\wp^2}{\hbar^2}
    \int_{V}d\vec{x} \left| 
      \tilde{D}^{\rm ret}(\vec{r}, \vec{x},\Delta,t) \right|^2
    \, P^{\rm (1s)} (\vec{x},\Delta,t) \nonumber \\
    &&+ \frac{\wp^2}{\hbar^2} \iint_Vd\vec{x}_1\vec{x}_2 
    \tilde{D}^{\rm  ret}(\vec{r},\vec{x}_1,\Dl,t) \nonumber \\
   && \qquad \tilde{D}^{*\rm ret}(\vec{r},\vec{x}_2,\Dl,t)
    \, P^{\rm (2s)}(\vec{x}_1,\vec{x}_2,\Dl,t)
    \nonumber \\
    \lefteqn{\Bar{\Gm}(\Dl,t) 
    \,=\,
    \frac{\wp^2}{\hbar^2}
    \tilde{D}^{-+}(\vec{r}_1\ne\vec{r}_2,\Dl,t) \;= } \\
    &=& 
    \frac{\wp^2}{\hbar^2} \int_{V}d\vec{x} 
    \tilde{D}^{\rm ret}(\vec{r}_1,\vec{x},\Dl,t)\nonumber \\
   && \qquad \tilde{D}^{*\rm ret}(\vec{r}_2,\vec{x},\Dl,t)
    \, P^{\rm (1s)}(\vec{x},\Dl,t) + \nonumber \\
    &&+ 
    \frac{\wp^2}{\hbar^2} \iint_Vd\vec{x}_1\vec{x}_2 
    \tilde{D}^{\rm  ret}(\vec{r}_1,\vec{x}_1,\Dl,t)\nonumber\\
   && \qquad \tilde{D}^{*\rm ret}(\vec{r}_2,\vec{x}_2,\Dl,t)
    \, P^{\rm (2s)}(\vec{x}_1,\vec{x}_2,\Dl,t)
    \nonumber
  \end{eqnarray}
\end{subequations}
where we use the Fourier transforms for all the functions.
The $\Dl'$ now denotes the detuning from resonance of the particular Fourier components at frequency $\omega=\omega_{ab}+\Dl'$, where $\omega_{ab}$ is the atomic resonance frequency.
In order to solve \eqs{eq:GammaGammabar} we make an additional approximation
where we neglect the coordinate dependence of the atomic variables $a,n,\rho_{ab,ba}$ on $\vec r_1, \vec r_2$ in the integration. This is justified if we assume a much weaker coordinate dependence of the atomic dynamics than of the field correlations.
In Ref.\,\cite{FleischYelin1999:lorentz} we have discussed in
great detail a way of calculating the 
kernel~$\tilde{D}^{\rm ret}(\vec{x}_1,\vec{x}_2,\Dl,t)$.
Here we just give the final result
\begin{equation}
  \label{eq:D}
  \tilde{D}^{\rm ret}(\vec{x}_1,\vec{x}_2,\Dl,t) = 
  -\frac{i\,  \hbar \, \omega^2}{6 \, \pi \, \epsilon_0 \, c^2} \,
  \frac{e^{-i \,q_0(\Delta) \, r }}{r},
\end{equation}
where $r=\left| \vec{x}_1 - \vec{x}_2 \right|$. Here, an average over all dipole directions is taken.
The wave vector $q_0$ is given by
\begin{equation}
  \label{eq:q0}
  q_0(\Delta)  = \frac{\omega}{c} \left(
    1 + i \, \frac{\hbar \, \omega}{3 \, \epsilon_0 \, c} \,
    \tilde{P}^{\rm (1)ret} (r, \Delta; t)
  \right).
\end{equation}
In \eqs{eq:D} and \noeq{eq:q0}, $\omega$ stands for the frequency of the emitted
light.

We then obtain a self-consistent expression for the quantum corrections as follows
\begin{subequations}
  \label{eq:gammabargamma1and2}
  \begin{eqnarray}
    \Gm &=& \gm \, \frac{a(t)}{2a(t)-1} \, 
    \left( \e^{2\zeta(\Dl)}-1 \right) \\
   && +
    2\gm\Cc^2\varrho^4 \, \frac{\gm\tGm}{\tGm^2+\Dl^2} \,
    \rho_{ab,ba}(t) \, I\left(\zeta(\Dl),\tilde{\varrho}(\Dl)\right),\nonumber \\ 
    \bar{\Gm}  &=& 3 \gm\Cc\varrho \, 
    \frac{\gm\tGm}{\tGm^2+\Dl^2} \, a(t) \, 
    I\left(\zeta(\Dl),\tilde{\varrho}(\Dl)\right) \\
    && + 2\gm\Cc^2\varrho^4 \, \frac{\gm\tGm}{\tGm^2+\Dl^2} \,
    \rho_{ab,ba}(t) \, I\left(\zeta(\Dl),\tilde{\varrho}(\Dl)\right).\nonumber
  \end{eqnarray}
\end{subequations}
In the expressions above both decay rates, $\Gm$ and $\bar\Gm$,
themselves implicitely depend on the atomic variables $a,n,d$, and $\rho_{ab,ba}$ 
of \eq{eq:2level:newvariables}, so they have to be self-consistently
calculated.
Different quantities that appear in \eq{eq:gammabargamma1and2} are given
as follows,
\begin{equation}
  \label{eq:params}
  \begin{array}{lclllcl}
    \tGm &=& \frac{\gm}{2} + \Gm,
    & \quad &
    \zeta(\Delta) 
    &=& \zeta_0 \, 
    \frac{\tilde{\Gm}^2}{\tilde{\Gm}^2+\Dl^2},\\
    \zeta_0 &=& \frac{1}{2} \, 
    \Cc\varrho\, \frac{\gm}{\tilde{\Gm}} \, (2a(t)-1),
    & \quad &
    \tilde{\varrho}(\Dl) &=& 
    \varrho - \frac{\Dl}{\tilde{\Gm}} \, \zeta(\Dl)\\
    \varrho &=& \pi\, \frac{d}{\lambda} \;=\; \frac{\om d}{2c},
    &&&& \\
    I(\zeta,\varrho) &=& \multicolumn{3}{l}{
    \frac{ 
      \left( \left(\zeta-1\right) \e^\zeta+\cos{\varrho} \right)^2 
      + \left(\varrho\e^\zeta-\sin{\varrho}\right)^2
    }{\left(\zeta^2+\varrho^2\right)^2}.}
  \end{array}
\end{equation}
We see that the \eqs{eq:gammabargamma1and2} and \noeq{eq:params} rely on
two parameters:
effective density, which is also known as the cooperativity parameter,
$\Cc$, given by 
\begin{equation}
  \label{eq:C}
  \Cc = \frac{2 \, \pi \, c^3 \, \Nc}{\omega^3},
\end{equation}
which is proportional to the total number of atoms in a cubic
wavelength of the emitted radiation, where $\Nc$ is
the density of the atoms;
and the effective radial size, $\varrho$, \eq{eq:params}, proportional
to the ratio of diameter $d$ of the sample to the wave length $\lambda$
of the emitted radiation.

Now, the role of $\Dl=2 H_{ii}/\hbar$ becomes clear:
As a real part of the dispersion, $\Dl$ is related to imaginery part
$\Gm$, a decay rate due to cooperativity effects, via a Kramers-Kronig
relationship
\begin{equation}
  \label{eq:delta:kk}
  \Dl = \frac{1}{\pi} {\cal P} 
  \int\limits_{-\infty}^\infty d\Dl' \, \frac{\Gm(a,x,\Gm,\Dl')}{\Dl-\Dl'}.
\end{equation}
In a two-level system, $\Dl$ is very small, and we can set it self-consistently to zero in what follows. The detailed format of $\Dl$ will be discussed in an upcoming publication.

\subsection{Cooperative Phenomena}

A physical measure of the dynamics in the system is the average intensity 
of the emitted radiation, rather than the effective population of
the excited state.
The intensity of radiation is directly proportional to the term
$- \dot a(\tau)$ by energy conservation.
One has to keep in mind that in our modeling we have neglected the
time retardation effects within the sample. 
When detecting the emitted radiation one would have to add some delay
before the radition reaches the detectors.

\begin{figure}[htbp]
  \centering
  \includegraphics[clip,width=0.85\linewidth]{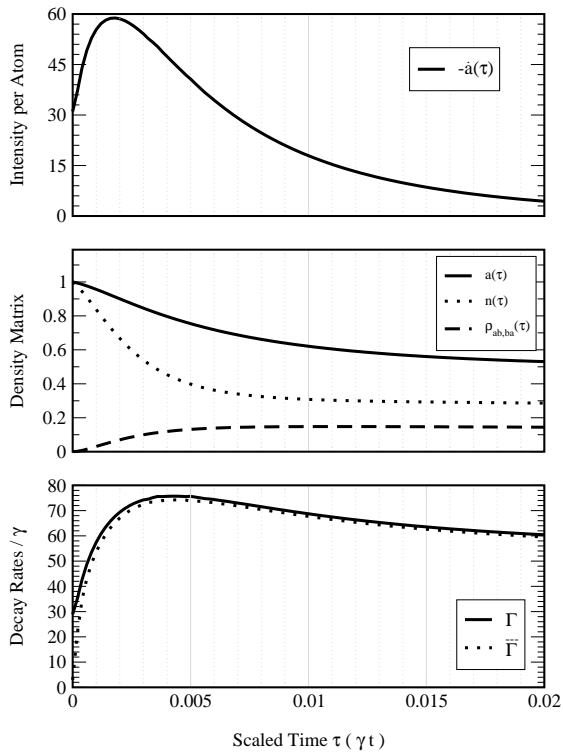}
  \caption{%
    Cooperative effects in optically dense medium at short 
    time scales ($\ll\gamma^{-1} \equiv 1$).
    (Top panel) Intensity per atom, $-\dot a(\tau)$, of spontaneous
    radiation is at the initial moments much stronger than that of a 
    single atom.
    (Middle panel) Behavior of the atomic variables shows an expected
    decay of the excited state, $a(\tau)$, and the population difference,
    $n(\tau)$), with a slow build-up of entanglement, $\rho_{ab,ba}(\tau)$, 
    from initial zero value.
    (Lower panel) Self-consistently calculated $\Gamma$ and $\bar\Gamma$
    along the trajectory of the system \noeq{eq:ss:sys} in phase space.
    Found values are two orders of magnitude larger than the spontaneous
    decay rate $\gamma$ (here $\gamma\equiv 1$).
    \label{fig:c=10:all}
    }
\end{figure}

We now examine short time behavior of the system \noeq{eq:ss:sys} when all
the atoms are initially in the excited state, \ket{a}. 
The values for the atomic variables
$a(0), \, x(0)$ and $n(0)$ are then $a(0) = 1, n(0) = 1, \rho_{ab,ba} = 0$.
While the system of equations \noeq{eq:ss:sys} provides a way of calculating
the chirp $\Dl$ from the spectral distribution $\Gm = \Gm(a,n,x,\Delta)$,
for simplicity we set $\Dl \equiv 0$ for rest of the discussed calculations.
We choose as values for the effective density $\Cc = 10$ and the
effective size $\varrho =10$.
Different aspects of the solution of system \noeq{eq:ss:sys},
given the initial values and relevant parameters as described above, 
is shown in \fig{fig:c=10:all}:
(Top panel) reveals a typical signature of the cooperative enhancement to the
single atom emision rate at the early stages of deexcitation. 
(Middle panel) depicts the quantities $\rho_{ab,ba}=\rho_{ab,ba}(\tau)$ 
and $n=n(\tau)$. 
Here, the most striking is the behavior of the two-atom entanglement, 
$\rho_{ab,ba}$, which grows from zero to some maximum value before dropping
back to zero (for times $\ll \gamma^{-1}$, not shown here).
This feature, a build-up of entanglement with a significant increase in
emitted intensity per atom compared to a single atom, is a
characterteristic signature of superradiance, as discussed in, e.g., 
Ref.~\cite{GrossHaroche1982}.

\begin{figure}[htbp]
  \centering
  \includegraphics[clip,width=0.85\linewidth]{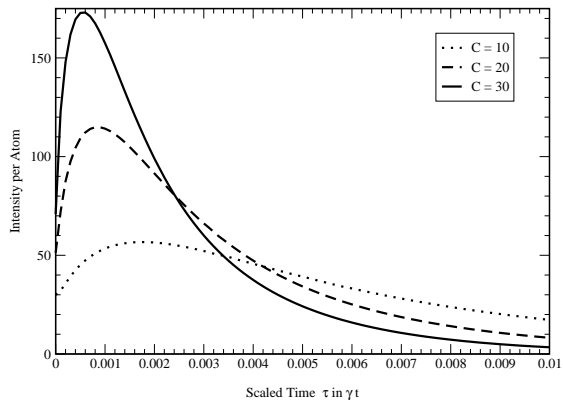}
  \caption{%
    Superradiant intensity per atom, $-\dot a(\tau)$, in dense optical
    media of cooperativity parameter $\Cc = 10, 20, 30$.
    Greater the cooperativity parameter, greater the maximum momentary 
    deexcitation rate $\Gamma = \Gamma(\tau)$, 
    see also \tab{tab:CvsmaxGamma}.
    \label{fig:cc}
    }
\end{figure}

One of the features of the superradiance is that the intensity per
atom depends strongly on the density of an optical medium under consideration.
In our theory the measure of density is the effective density
$\Cc$, \eq{eq:C}. We perform a series of calculations where we vary
$\Cc$, and calculate
$\max\{ -\dot a(\tau) \}_{\tau} = -\dot a(\tau_{\max})$
and $\tau_{\max}$ for each of the solutions.
The results of this calculation are shown in \tab{tab:CvsmaxGamma}:
maxima of the decay rate are proportional to the effective density$\Cc$, 
while the times at which maxima occur depend are reversely proportional
to the effective density $\Cc$.
This is in good agreement with the expected superradiance scaling,
that is, the emission per atom $-\dot a(\tau)$ is indeed proportional to $\Cc$,
while the times at which the maxima are obtained scale like $1/\Cc$.
\begin{table}[htbp]
  \centering  
  \begin{tabular}{|c|cc|}
    \hline
    $\Cc$ & $\max\{ -\dot a \}_{\tau}$ 
    & $\tau_{max}$ (in $\gamma^{-1}$)\\
    \hline
    10 &  57 & 0.0018\\
    20 & 115 & 0.0008\\
    30 & 172 & 0.0006\\
    \hline
  \end{tabular}
  \caption{%
    Dependence of the maximum of the intensity per
    atom $-\dot a$ and the scaled time $\tau_{\max}$, at which
    the maximum occured, as a function of $\Cc$.
    The quantity $-\dot a$ is a linear function of $\Cc$,
    and thus confirming the superradiant nature of the decay.
    The times $\tau_{\max}$ at which the maximum of the radiation is
    achieved, scale like $1/\Cc$.
    \label{tab:CvsmaxGamma}
  }
\end{table}
The curves with the calculated intensity per single atom for the cooperativity
parameter $\Cc = 10,20,30$ are given in \fig{fig:cc}.

\begin{figure}[htbp]
  \centering
  \includegraphics[clip,width=.85\linewidth]{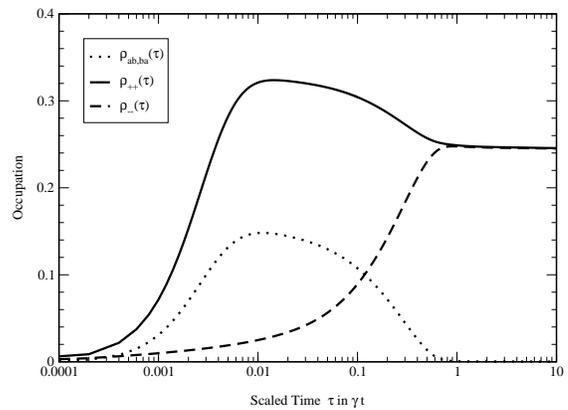}
  \caption{%
    Superradiant, $\rho_{++}$, and subradiant $\rho_{--}$ density matrix
    element in logarithmic time for a dense optical media with $\Cc = 10$.
    The solution of system \noeq{eq:ss:sys} shows that 
    in a superradiant decay both, superradiant and subradiant, states
    get populated.
    As expected, there is a significant difference between the two, as long
    as there is an entanglement present, $\rho_{ab,ba}(t)\neq0$.
    \label{fig:supersub}
    }
\end{figure}

In the original Dicke's paper~\cite{Dicke1953}, the superradiance occurs
through a decay over the subset of symmetric many-atom states. 
In our effective two-atom two-level model, 
we may introduce two alternative matrix elements: a subradiant, $\rho_{--}$,
for the population of anti-symmetric two-atom state, 
and a superradiant, $\rho_{++}$, for the symmetric two-atom state,
where $\ket{\pm} = 1/\sqrt2(\ket{ab} \pm \ket{ba})$. 
This is permissible if the size of the atomic sample is small, because
the effects of change of phase due to propagation can be neglected.
In terms of atomic variables $a(\tau), n(\tau)$ and $\rho_{ab,ba}(\tau)$, these
matrix elements are 
$ \rho_{\pm\pm} = 1/4 ( 1 - n(\tau) \pm 4\, \rho_{ab,ba}(\tau) )$,
The behavior of the matrix elements under the conditions exhibiting
superradiant behavior, \eq{eq:ss:sys}, and with the cooperativity
parameter $\Cc = 10$ is shown in \fig{fig:supersub}.
Here we observe that the dipole-dipole interaction, which is built into
our model, indeed contributes to populating the anti-symmetric two-atom
state, whereas the population of $\ket{+}$ and $\ket{-}$ equalizes for long times when decoherence overrules correlation. This feature is absent in standard treatments of 
superradiation, e.g. \cite{GrossHaroche1982}.

\section{Correlation and entanglement}
In this paragraph we would like to clarify the respective roles of crrelation and entanglement in superradiance. Correlation, i.e., a two-atom cooperative effect, is stored in the non-diagonal two-atom matrix element, $\rho_{ab,ba}$, and can be measured as the energy splitting between the $\ket{+}$ and $\ket{-}$ state. This quantity clearly is responsible for the build-up of superradiation (see, e.g., Eqs.~(\ref{eq:2level:sys}), (\ref{eq:ss:sys}), and (\ref{eq:gammabargamma1and2})). Entanglement, however, defined to be a quantity describing how far the system is away from a product state $\rho_{\rm prod}=\sum_i\rho_i^{(1)}\otimes\rho_i^{(2)}$, where $\rho_i^{(1/2)}$ are the single atom density matrices. Wootters \cite{wootters} has found a measure for the entanglement of two systems. It turns out that in all two-atom superradiant systems, even those with a maximum grade of coherence, there is no entanglement present. The proof is easy, since $\rho_{abba}$ always has to be smaller or equal to $\sqrt{\rho_{aa}\rho_{bb}}$, where $\rho_{aa}$ and $\rho_{bb}$ are the single-atom populations.

Obviously, at this time it is not possible to make any final statement about the entanglement of more than two atoms with cooperative interactions, but since we treat only effects up to second order in the atomic interaction exactly, it is quite clear that also in our system there is no entanglement present. In conclusion, superradiance is an effect based on correlation, but not on entanglement.
\section{Conclusion}

We have presented a method for modeling the cooperative effects in a
dense optical medium, based on an effective two-atom two-level model.
In its simplest version, a small sample approximation, we are able to
obtain a system of equations which describes the evolution of a two-atom
two-level density matrix.
Solving the system for an initially inverted system, 
we are able to predict the cooperative behavior with all
the features of superradiance:
coherent build-up of entanglement, 
with the decay rate up to two orders of magnitude greater than the
spontaneous decay rate.
Additionally, the maximum of the decay rate $\Gamma$ 
scales linearly with the number of particles indicating that the maximum
intensity of the deexcitation pulse scales as $N^2$, the square of the
number of particles.  
Our model requires two parameters, the cooperativity parameter $\Cc$,
and the effective size of the system in terms of the emitted wavelength.
Theory predicts two novel phenomena: chirp, or the change in transition
frequency due to the cooperative phenomena and subradiance, both of which are outside
a standard description of superradiance.
Accounting for the shape of the optical medium in the formalism, 
and in particular analysis of cigar-shaped atomic sample, 
will be subject of future work.

\bibliography{%
  bibliography/superradiance,%
  bibliography/bec,%
  bibliography/numericks,%
  bibliography/generalphysics,%
  bibliography/lightstorage%
}
 
\end{document}